\documentclass[mathleft]{an}
\usepackage{natbib}
\usepackage{graphicx}
\usepackage{times}
\begin{document}

\Pagespan{789}{}
\Yearpublication{2007}%
\Yearsubmission{2007}%
\Month{12}%
\Volume{999}%
\Issue{88}%

%
   \title{A new period determination for the close PG1159 binary
   \smallskip\linebreak\mbox{SDSS~J212531.92$-$010745.9}}


   \author{S. Schuh\inst{1}\fnmsep\thanks{Corresponding author:
       \email{schuh@astro.physik.uni-goettingen.de}}
          \and
          I.~Traulsen\inst{1}
          \and
          T.~Nagel\inst{2}
          \and
          E.~Reiff\inst{2}
          \and
          D.~Homeier\inst{1}
          \and
          H.~Schwager\inst{2}
          \and
          D.-J.~Kusterer\inst{2}
          \and
          R.~Lutz\inst{1}
          \and
          M.~R.~Schreiber\inst{3}
          }

   \institute{Institut f\"ur Astrophysik, Georg-August-Universit\"at G\"ottingen,
              Friedrich-Hund-Platz~1, 37077 G\"ottingen, Germany
         \and Institut f\"ur Astronomie und Astrophysik, Eberhard-Karls-Universit\"at
              T\"ubingen, Sand~1, 72076 T\"ubingen,
	      Germany
         \and Departamento de Fisica y Meteorologia, Facultad de
              Ciencias, Universidad de Valparaiso, Valparaiso, Chile 
   }

   \received{30 Aug 2007}
   \accepted{later}
   \publonline{later}


  \abstract{%
Methods to measure masses of PG\,1159 stars in order
    to test evolutionary
    scenarios are currently based on spectroscopic masses or
    asteroseismological mass determinations. One recently discovered
    PG\,1159 star in a close binary system may finally allow the first
    dynamical mass determination, which has so far been analysed on
    the basis of one SDSS spectrum and photometric monitoring.
\smallskip\newline
In order to be able to phase future radial velocity measurements
    of the system \mbox{SDSS~J212531.92$-$010745.9}, we follow up on
    the photometric monitoring of this system to provide a solid
    observational basis for an improved orbital ephemeris determination.
\smallskip\newline
New white-light time series of the brightness variation of
    \mbox{SDSS~J212531.92$-$010745.9} with the T\"ubingen 80\,cm and
    G\"ottingen 50\,cm telescopes extend the monitoring into a second
    season (2006), tripling the length of overall observational data
    available, and significantly increasing the time base covered.
\smallskip\newline
We give the ephemeris for the orbital motion of the system, based on a sine fit
    which now results in a period of 6.95573(5)\,h, and discuss the associated
    new amplitude determination in the context of the phased light
    curve variation profile. 
    The accuracy of the ephemeris has been improved by more than one order of
    magnitude compared to that previously published for 2005 alone.
  }

   \maketitle
%
%
\section{Introduction}
PG\,1159 stars are hot hydrogen-deficient pre-WDs which are believed
to be the outcome of a late or very late helium-shell flash 
during their post-AGB evolution. About 40 such objects are 
known at present \citep{2006PASP..118..183W}, and a
subset of currently 11 objects forms the class of the pulsating GW~Vir
variables. From their evolutionary history, typical masses should be
around 0.6\,M$_{\sun}$; spectroscopic and asteroseismological mass
determinations confirm this. These methods however rely on stellar
structure and evolution modelling, and can so far not be tested
independently.
\par
\citet{2006A&A...448L..25N} have recently announced the unique discovery of a
PG\,1159 star in a close binary system, which might for the first time
allow a dynamical mass determination for a PG\,1159 star. 
\object{SDSS~J212531.92-010745.9} was discovered to show H$\alpha$
emission during a systematic search of SDSS archival data for 
white dwarf + main sequence 
companion candidates. It was subsequently classified as a PG\,1159
star from a spectral analysis of the SDSS spectrum, and found to
undergo brightness variations in follow-up time-resolved
photometry. The light curve shows a periodicity of 6.95616(33)\,h, 
attributable to orbital motion, with a flat bottom part, and no eclipses.
The periodic brightening with a peak-to-peak amplitude of 0.7\,mag 
can be interpreted as the light contribution by the irradiated side of
the cool companion.
\par
We have continued the photometric coverage of this system during the
following 2006 observing season, and report the improved ephemeris
with a more accurate period determination in this paper.
%
%
\section{Observations}
Following up the light curve of \object{SDSS~J212531.92-010745.9}
presented by \citet{2006A&A...448L..25N}, we have obtained new
observations with the T\"ubingen 80\,cm and the G\"ottingen 50\,cm
telescopes during 19 nights between September and November 2006. The full data
set now available is listed in Table~\ref{tab_obs} with details of the set-up.
\par
Data reduction was done using the TRIPP routines
\linebreak\citep{2003BaltA..12..167S}.
The images were bias, dark, and flat field corrected, and relative
light curves were obtained
from aperture photometry using the same set
of reference stars for all data sets. 
The flux of the combined data was normalised to its mean value, and
all times were heliocentrically corrected.
%
\begin{figure*}
  \centering
  \includegraphics[width=1.\textwidth]{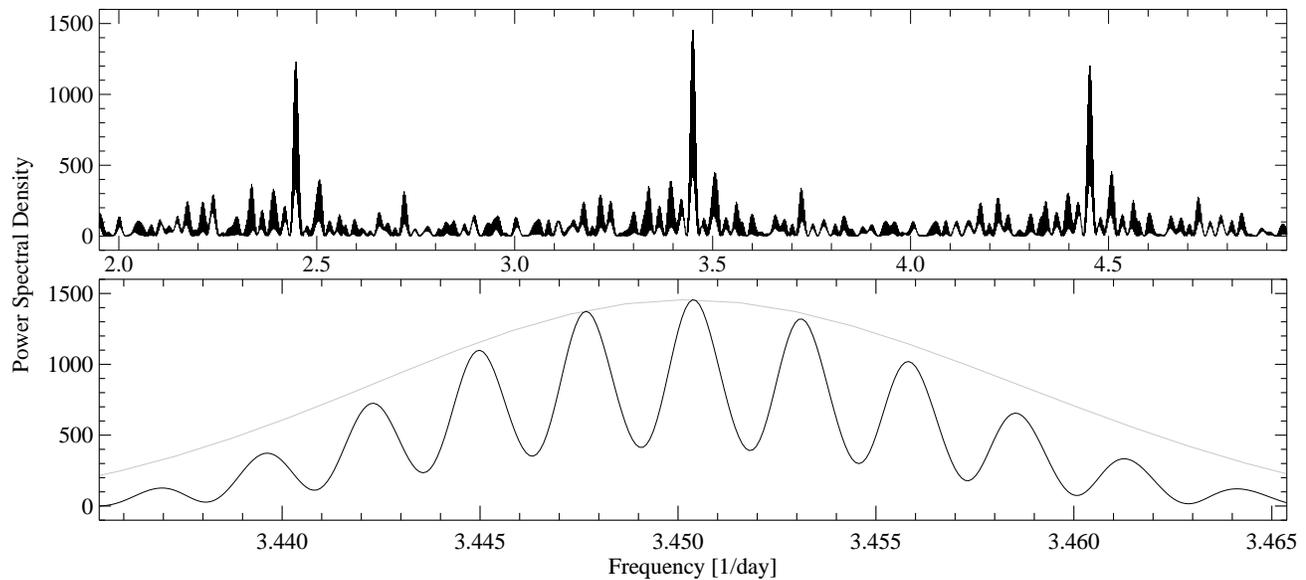}
  \caption{Periodogram of the combined light curve:
  frequency range illustrating the one-day (upper panel) and
  the yearly alias patterns (lower panel),
  compared to the central peak of the periodogram obtained from the 2005
  data alone (grey line).}
  \label{psd}
\end{figure*}
\begin{figure*}
  \centering
  \includegraphics[width=.65\textwidth]{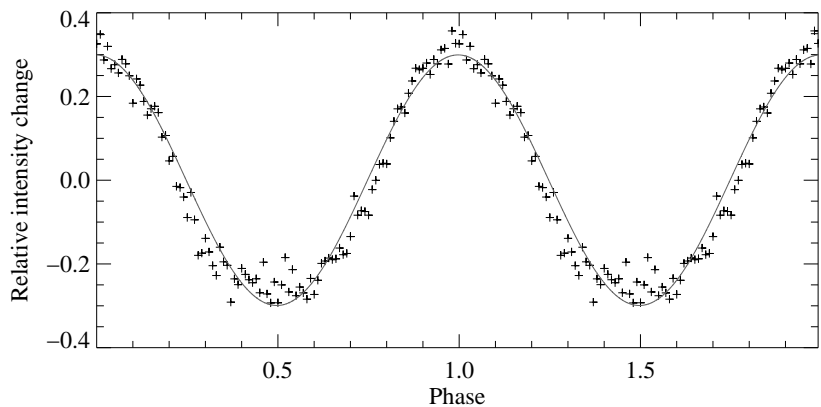}
  \caption{Profile of the combined light curve obtained by folding it
  onto the orbital period and rebinning into 100 phase bins (crosses),
  and overplotted sine fit (grey line).} 
  \label{profile}
\end{figure*}
\begin{table*}
\centering
\caption{Observation log. All observations were performed with clear
  filter.}
\begin{tabular}{lrrrcl}
  \hline
  \noalign{\smallskip}
  Date  &  t$_{\rm exp}[s]$&  t$_{\rm cycle}[s]$  &  Duration$[s]$  &  Telescope & Camera\\
  \noalign{\smallskip}
  \hline
  \noalign{\smallskip}
  2005/09/21 &  90 &  98 & 18900 & 80\,cm & ST-7E\\
  2005/09/22 &  90 &  98 & 18899 & 80\,cm & ST-7E\\
  2005/09/23 &  90 &  98 & 21758 & 80\,cm & ST-7E\\
  2005/09/23 & 180 & 194 &  4051 & 50\,cm & STL-6303E\\
  2005/09/23 & 240 & 254 &  9656 & 50\,cm & STL-6303E\\
  2005/10/06 & 240 & 248 & 10202 & 50\,cm & STL-6303E\\
  2005/10/07 & 240 & 246 & 14897 & 50\,cm & STL-6303E\\
  2005/10/08 & 240 & 248 &  9298 & 50\,cm & STL-6303E\\
  2005/10/10 &  90 &  98 & 19852 & 80\,cm & ST-7E\\
  2005/10/11 & 240 & 248 & 17872 & 50\,cm & STL-6303E\\
  2005/10/18 &  90 &  98 & 16532 & 80\,cm & ST-7E\\
  2005/10/26 &  90 &  98 & 20095 & 80\,cm & ST-7E\\
  2006/09/12 &  60 &  63 & 15919 & 80\,cm & STL-1001E\\
  2006/09/13 &  60 &  63 & 21312 & 80\,cm & STL-1001E\\
  2006/09/20 &  60 &  63 & 23056 & 80\,cm & STL-1001E\\
  2006/09/20 & 240 & 247 & 17321 & 50\,cm & STL-6303E\\
  2006/09/21 &  60 &  63 & 22489 & 80\,cm & STL-1001E\\
  2006/09/21 & 180 & 187 & 21441 & 50\,cm & STL-6303E\\
  2006/09/22 &  60 &  63 & 22299 & 80\,cm & STL-1001E\\
  2006/09/22 & 180 & 187 &  1922 & 50\,cm & STL-6303E\\
  2006/09/22 & 420 & 427 & 13388 & 50\,cm & STL-6303E\\
  2006/09/23 &  90 &  93 & 22008 & 80\,cm & STL-1001E\\
  2006/09/23 & 180 & 187 & 20595 & 50\,cm & STL-6303E\\
  2006/09/24 &  60 &  63 &  9805 & 80\,cm & STL-1001E\\
  2006/09/24 & 240 & 247 & 24678 & 50\,cm & STL-6303E\\
  2006/09/27 &  90 &  93 &  9470 & 80\,cm & STL-1001E\\
  2006/09/27 & 240 & 247 &  3205 & 50\,cm & STL-6303E\\
  2006/10/08 & 240 & 247 & 11601 & 50\,cm & STL-6303E\\
  2006/10/09 & 240 & 247 & 19808 & 50\,cm & STL-6303E\\
  2006/10/10 & 240 & 247 & 19241 & 50\,cm & STL-6303E\\
  2006/10/11 & 240 & 247 & 19721 & 50\,cm & STL-6303E\\
  2006/10/12 &  60 &  63 & 18105 & 80\,cm & STL-1001E\\
  2006/10/16 &  60 &  63 & 17602 & 80\,cm & STL-1001E\\
  2006/10/16 & 240 & 247 & 16763 & 50\,cm & STL-6303E\\
  2006/10/17 &  60 &  63 & 13798 & 80\,cm & STL-1001E\\
  2006/10/17 & 120 & 123 &  4052 & 80\,cm & STL-1001E\\
  2006/10/17 & 240 & 247 & 18758 & 50\,cm & STL-6303E\\
  2006/10/26 &  60 &  63 & 16031 & 80\,cm & STL-1001E\\
  2006/10/27 &  60 &  63 & 19739 & 80\,cm & STL-1001E\\
  2006/10/30 &  40 &  43 & 11609 & 80\,cm & STL-1001E\\
  2006/11/15 & 240 & 247 & 18264 & 50\,cm & STL-6303E\\
  \noalign{\smallskip}
  \hline
\end{tabular}\label{tab_obs}
\end{table*}
%
%
%
%
\section{Results}
In Fig.\,\ref{psd}, we show the periodogram of the combined light
curve from 2005 and 2006. The correct position of the dominant frequency can
unambiguously be identified from the strong alias pattern. The gain in accuracy
over the 2005 data alone is illustrated by the central peak (grey line in the
lower panel).  
By assuming the corresponding period as a start value, the combined light
curve was fitted with a non-linear least-squares sine fit. It resulted in
an improved period determination of 6.95573(5)\,h and a sinusoidal
amplitude of 0.299(33) relative intensity change (0.284(28)\,mag).
The amplitude (and its uncertainty, determined as the 1\,$\sigma$ residual
scatter of the original data after subtraction of the mean profile)
should be considered as an estimate of the magnitude of the
brightness change only (for details see the Discussion section).
\par
Referring to the most recent observation to define the zero point, we
determine the ephemeris of predicted 
\emph{maxima} times to be
\begin{center}
\[ 
\begin{array}{lr@{}lcr@{}l}
 &  {\rm HJD} = 2454055\fd 213&4(4)  & + &    0\fd 28982&2(2) \cdot E. \\
\end{array} 
\]
\end{center} 
The uncertainty in the period corresponds to the 1\,$\sigma$ formal
fitting error, while the uncertainty in the zero point corresponds to
half the cycle time in the short-exposure runs.
%
%
%
\section{Discussion}

In Fig.\,\ref{profile}, we compare the folded profile to the sinusoidal
amplitude of 0.299(33) relative intensity change. 
Clearly, the observed peak-to-peak variation is not fully reproduced
by the fit, owing to the broader trough and the narrow maximum already
described  by \citet{2006A&A...448L..25N}, who fitted magnitude changes, while
we now fit intensity changes. For a higher fit amplitude
and/or raised zero line, the observed raising and falling edges would
lie significantly below the sine fit flanks.
Although the light curve solution presented by
\citet{2006A&A...448L..25N} still suffers from a degeneracy in mass
and radius for the PG\,1159 component, this light curve shape
emphasizes the superiority of their more realistic model over a plain
sine fit and validates their interpretation of the observed
brightening as a reflection effect. 
\par
The refined ephemeris and period determination will allow a precise
extrapolation of the ephemeris well into the next observing seasons, and
facilitate the phasing of radial velocity measurements. This
additional information will help to lift the existing degeneracy, so
that we will be able to present a more elaborate light curve model in
a future paper, and derive the masses in the system. 
%
\acknowledgements
We thank
Sylvia Brandert,
Sebastian Wende,
Jens Adamczak,
Agnes Hoffmann, 
Ralf Kotulla,
Simon H\"ugel\-meyer,
Markus Hundertmark,
and Johannes Fleig for supporting the 2006 observations;
and Klaus Reinsch for encouragement and technical support.
We gratefully acknowledge the initialization of this project by
Boris G\"ansicke who first directed our attention to this special
object. 
\end{document}